\documentclass[12pt]{article}
\usepackage{amsmath}
\usepackage{bm}
\usepackage{bbm}
\usepackage{cite}
\usepackage[dvips]{color}

\usepackage{graphicx}
\usepackage{slashed}
\usepackage{tabularx}
\usepackage{comment}
\usepackage{hyperref}

\begin{document}
\title{
\vspace{-1.2truecm}
\Large\bf
 Performance of new 8-inch photomultiplier tube used for the Tibet muon-detector array
}
\author{
Ying Zhang$^a$$^b$, Jing Huang$^a$, Ding Chen$^c$,\\ Liu-Ming Zhai$^c$,
Xu Chen$^a$,Xiao-Bin Hu$^a$$^d$, \\Yu-Hui Lin$^a$, Hong-Bo Jin$^c$,
Xue-Yao Zhang$^d$,\\ Cun-Feng Feng$^d$, Huan-Yu Jia$^e$, Xun-Xiu Zhou$^e$,\\
DANZENGLUOBU$^f$, Tian-Lu Chen$^f$, LABACIREN$^f$ \\
Mao-Yuan Liu$^f$, Qi Gao$^f$ and ZHAXICIREN$^f$\footnote{Corresponded author Email: yingzhang@ihep.ac.cn}\\
  \small\textit{$^a$Key Laboratory of Particle Astrophysics,Institute of High Energy Physics,}\\
  \small\textit{Chinese Academy of Sciences,Beijing 100049, China}\\
  \small\textit{$^b$University of Chinese Academy of Sciences, Beijing, 100049, China}\\
  \small\textit{$^c$National Astronomical Observatories, Chinese Academy of Sciences, Beijing 100012, China}\\
  \small\textit{$^d$Department of Physics, Shandong University, Jinan 250100, China}\\
  \small\textit{$^e$Institute of Modern Physics, Southwest Jiaotong University, Chengdu 610031, China}\\
  \small\textit{$^f$Physics Department of Science School, Tibet University, Lhasa 850000, China}\\
}
\date{}
\maketitle
\begin{abstract}
A new hybrid experiment has been constructed to measure the chemical composition of cosmic rays around the ``knee"
in the wide energy range by the Tibet AS$\gamma$ collaboration at Tibet, China,  since 2014.
They consist of a high-energy air-shower-core array (YAC-II), a high-density air-shower array (Tibet-III) and
a large underground water-Cherenkov muon-detector array (MD). In order to obtain the primary proton, helium and iron spectra and
their ``knee" positions in the energy range lower than $10^{16}$ eV, each of PMTs equipped to the MD cell is required to measure the number
of photons capable of covering a wide dynamic range of 100 - $10^{6}$ photoelectrons (PEs) according to Monte Carlo simulations.
In this paper, we firstly compare the characteristic features between R5912-PMT made by Japan Hamamatsu and CR365-PMT made by Beijing Hamamatsu.
This is the first comparison between R5912-PMT and CR365-PMT. If there exists no serious difference, we will then add two 8-inch-in-diameter PMTs
to meet our requirements in each MD cell, which are responsible for the range of 100 - 10000 PEs and 2000 - 1000000 PEs, respectively.
That is, MD cell is expected to be able to measure the number of muons over 6 orders of magnitude.
\end{abstract}
\section{Introduction}
The energy spectrum of cosmic rays is well described by a
power law over a wide energy range covering more than 10
decades; this is regarded as a remarkable feature of the nonthermal acceleration mechanism of high-energy cosmic rays.
The power index suddenly steepens from approximately -2.7
to -3.1 at around 4$\times$$10^{15}$ eV, resulting in a distinctive ``knee"
shape in the spectrum\cite{1}\cite{2}. The origin of the ``knee" is
an outstanding problem in astroparticle physics\cite{3}\cite{4}. A lot of models
have been proposed such as a change of acceleration mechanisms
at the sources of cosmic rays (supernova remnants, pulsars, etc.), to a single-source assumption
 or effects due to propagation inside the galaxy
(diffusion, drift, escape from the Galaxy), and some unknown new processes in
the atmosphere during air-shower development\cite{5}\cite{6}\cite{7}\cite{8}\cite{9}.
Hence, in order to resolve the origin of the knee, it is needed to observe the break point of the spectral
index for individual chemical component\cite{3}\cite{10}\cite{11}.

In 2014, an important upgrade of the Tibet AS$\gamma$ project was
completed. They consist of an air-shower-core array (YAC-II),
a high-density air-shower array (Tibet-III) and a large underground water-Cherenkov muon-detector array (MD)\cite{12}\cite{13}.
For an air shower event, the Tibet-III provides the arrival direction ($\theta$) and the air shower size ($N_{e}$) which are
interrelated to primary energy, the YAC-II measures the high
energy electromagnetic particles in the core region so as to obtain the characteristic parameters of air-shower cores, at the same
time, the underground MDs record the high-energy muons above
1 GeV. The MD array now consists of 5 pools set up 2.5 meters underground,
each with 16 cells, covering a total area of $~$4500 $m^{2}$. Each cell of the MD array is composed of a concrete
water tank with a cubic form of 7.2 m wide, 7.2 m long, 1.5 m deep and two 20-inch PMTs (R3600)
are mounted downward on the ceiling. The method for obtaining the light-component
spectrum of primary cosmic rays at the ``knee" energies with the Tibet experiment is described
in the paper\cite{12}\cite{14}. In order to explicitly observe the break point of the spectrum
for individual chemical component, we need to observe primary particles with energy up to $10^{16}$ eV.
According to Monte Carlo simulation, a set of two PMTs in each MD cell is required to measure the number of
photons over a wide dynamic range from 100 to $10^{6}$ photoelectrons (PEs), as shown in Fig.1.

In this paper, we firstly compare the characteristic features between R5912-PMT made by Japan Hamamatsu and CR365-PMT made by Beijing Hamamatsu.
This is the first comparison between R5912-PMT and CR365-PMT. Since it is the first time for Beijing Hamamatsu
to produce this type of 8-inch-in-diameter PMTs, so it is needed to test the performance of PMTs whether they
are intend to satisfy our request, also to provide valuable feedback to the company producing them.
Then, we use a set of two PMTs and measure the linearity of the PMT response as a function of the light intensity
to meet our requirements.

\begin{figure}
\begin{center}
 \includegraphics[width=0.7\textwidth]{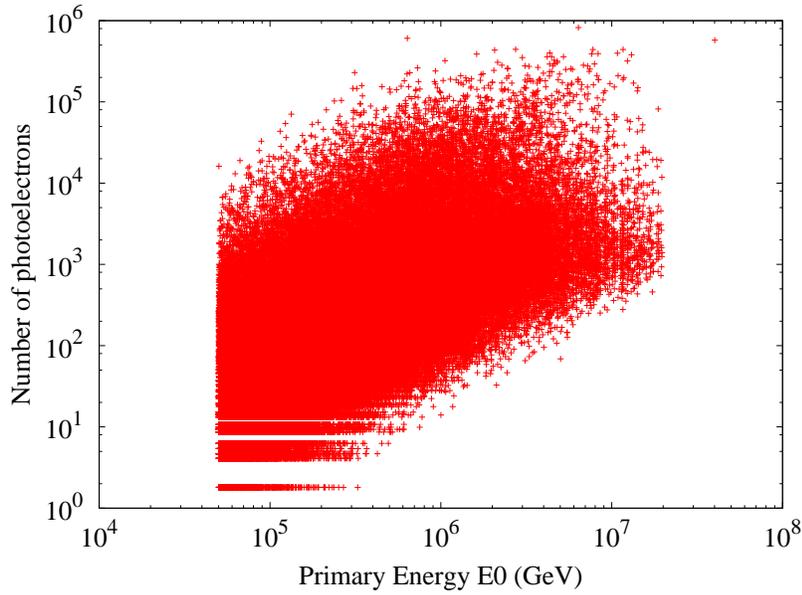}
\caption{\label{fig1} The correlation between primary energy and number of photoelectrons(PEs).}
\end{center}
\end{figure}

\section{The experimental setup}
In order to study the performance of these PMTs, the following quantities have been measured:

1)photoelectron spectra ,

2)gain as a function of high voltage,

3)dark count rate,

4)linearity of PMT output signals,

5)dynamic range of PMT.

The entire test system is mainly composed of Pulse Generator (Agilent Technologies 81160A), Laser (PiL044SM-SN-513B),
Filter, High Voltage power supply (CAEN N1470), Gate Generator (ORTEC GG8020),
Low Threshold Discriminator (CAEN N845), Scaler (ORTEC 772 counter), ADC (Lecroy 2249A, the resolution is 0.25 pC/count),
and data acquisition (DAQ) computer.
The test system is schematically shown in Fig. 2. The light source used for this test was a pico second laser
(Pil044SM-SN-513B) which has very narrow pulse width and high stability. Signal frequency applied to
the laser is controlled by the pulse generator (fixed)
and the light intensity is controlled by the control box (variable). At the same time, the pulse generator generates
the NIM signals to use as triggers. Before reaching the PMT, the light
passed through neutral density attenuation filters. With different filters,
we were able to change the light intensity by more than 5 orders of magnitude.
When we measure 1),2), 4)and 5), the signals from the PMT are then fed into a charge-integrating ADC in a
camac crate where they are measured and then read out by the DAQ computer;
when we measure 3), the signals from the PMT are then fed into a low threshold discriminator and then into a scaler.
All the analysis is then done on the computer.

\begin{figure}
\begin{center}
 \includegraphics[width=0.7\textwidth]{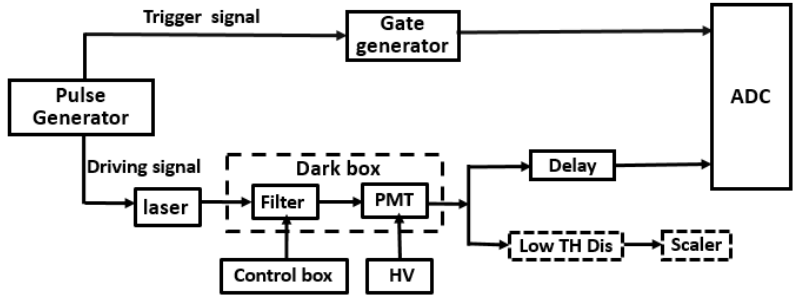}
\caption{\label{fig2} Schematic view of the test system.
When we measure 1),2), 4)and 5), the signals from the PMT are then fed into a charge-integrating ADC in a
camac crate where they are measured and then read out by the DAQ computer;
when we measure 3), the signals from the PMT are then fed into a low threshold discriminator and then into a scaler.}
\end{center}
\end{figure}

\section{Results}
Using this test system, the single photoelectron (SPE) spectra, multi photoelectrons spectra,
gain as a function of voltage, dark count rate at 1/3 photoelectron (PE) threshold and linearity of PMT output signals are measured for CR365
and R5912. Details of each test are described in the following sections.

\subsection{Single photoelectron spectra}
In order to get the absolute gain of the PMT at a certain voltage, the SPE
spectrum is measured. Fig. 3 shows the SPE spectra for
CR365 at 1500 V and R5912 at 1350 V corresponding to $\sim$1.5$\times$$10^{7}$ gain.
The first peak is the pedestal and the second peak is due to 1 photoelectron events.
To quantify the resolution of the single photoelectron spectrum, the peak to
valley ratio is used. The peak to valley ratio is the ratio of the maximum value
of the histogram of the SPE spectrum to the minimum value
between the pedestal and the maximum.
The SPE peak to valley ratio for CR365 and R5912 are both close to 2 which indicate that both PMTs
have a good charge resolution for detecting single photoelectron.
\begin{figure}
\begin{center}
 \includegraphics[width=0.7\textwidth]{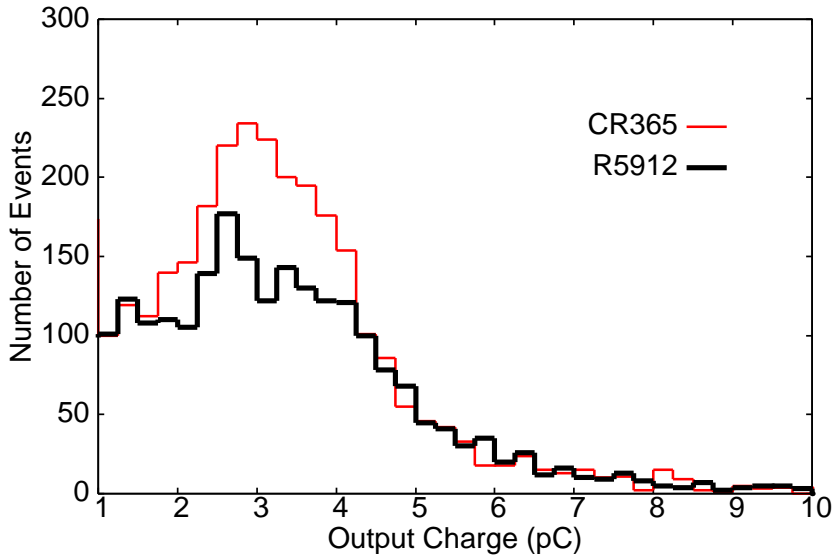}
\caption{\label{fig3} Single photoelectron (SPE) spectra at $\sim$1.5$\times$$10^{7}$ gain measured for CR365 and R5912.
The SPE peak to valley ratio for CR365 and R5912 are both close to 2 which indicate that both PMTs
have a good charge resolution for detecting single photoelectron.}
\end{center}
\end{figure}
\subsection{Multi photoelectron spectra}
We also measured multiple photoelectrons (multi-PEs) spectra for CR365 and R5912 both at 1800V
as shown in Fig. 4 and Fig. 5. Here, the resolution of ADC is 0.25 pC/count. A multi-gaussian fit was carried out in equation (1),
$N_{pe}$ is the average multi-PEs number,$\sigma$ is standard deviation of multi-PEs,
C is the ADC count of multi-PEs\cite{15}. From Fig. 4 and Fig. 5, we find the multi-PEs spectra of CR365 and R5912
can be well reproduced by this multi-gaussian function. It means that both PMTs
have a good charge resolution for detecting multi photoelectrons.
\begin{equation}\label{1}
  f(x)=\sum \frac{(N_{pe}^{n})(e^{-N_{pe}}}{n!})(2n\pi\sigma)^{2})^{1/2}
  e^{-(x-nC)^{2}/(2n\sigma^{2})}
\end{equation} 
 
\begin{figure}
\begin{center}
 \includegraphics[width=0.7\textwidth]{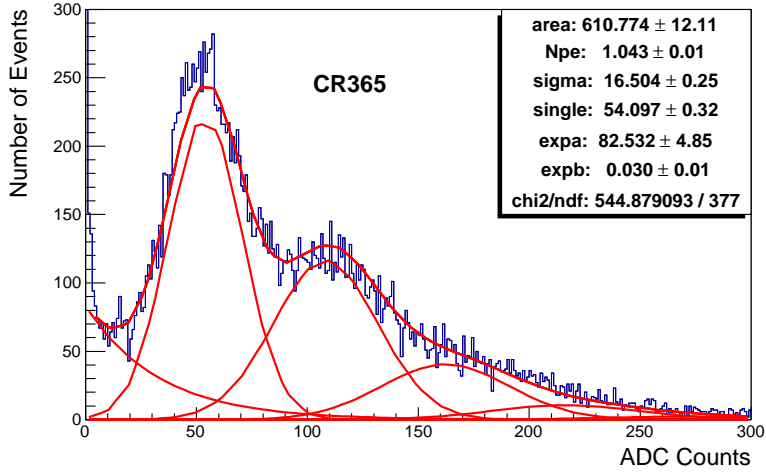}
\caption{\label{fig4} The response of CR365 versus number of photoelectrons shows
a good charge resolution for detecting multi photoelectrons. The blue points show the experimental data.
The red solid lines are fitting lines.}
\end{center}
\end{figure}
\begin{figure}
\begin{center}
 \includegraphics[width=0.7\textwidth]{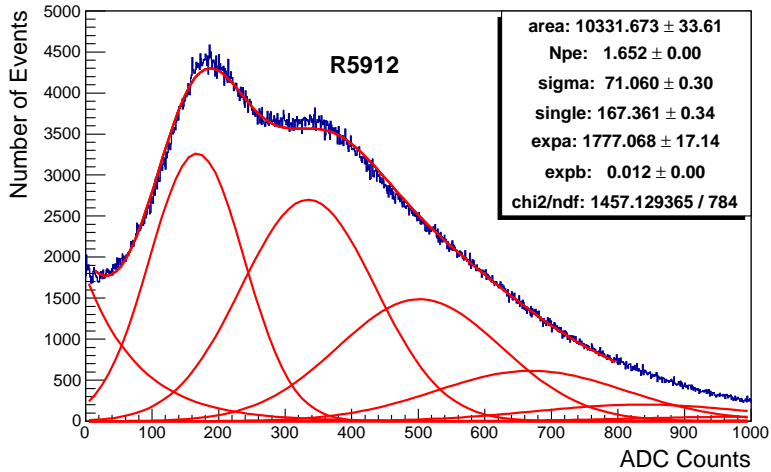}
\caption{\label{fig5} The multiple photoelectrons spectrum of R5912 also shows
a good charge resolution for detecting multi photoelectrons. The blue points show the experimental data.
The red solid lines are fitting lines.}
\end{center}
\end{figure}
\subsection{Gain as a function of high voltage}

An important feature is the gain curve with respect to the supplied high voltage (HV).
To get the gain as a function of high voltage, the laser light intensity has been fixed while the PMT HV has been
varied. The relationship between gain (G) and input high voltage (V) is described by
\begin{equation}\label{2}
  G=\alpha V^{\beta}
\end{equation}
with $\alpha$ and $\beta$ to be determined from measurements.
 The fit parameters are $\beta$=8.96$\pm$0.10 for R5912,
$\beta$=8.27$\pm$0.06 for CR365, respectively.
The absolute gain for each PMT was measured at a known voltage from the SPE
spectrum, and this can be used to convert this relative gain measurement into an
absolute one. Fig. 6 shows the absolute gain for CR365 and R5912 which is about three times of that CR365 at 1800V.
\begin{figure}
\begin{center}
 \includegraphics[width=0.7\textwidth]{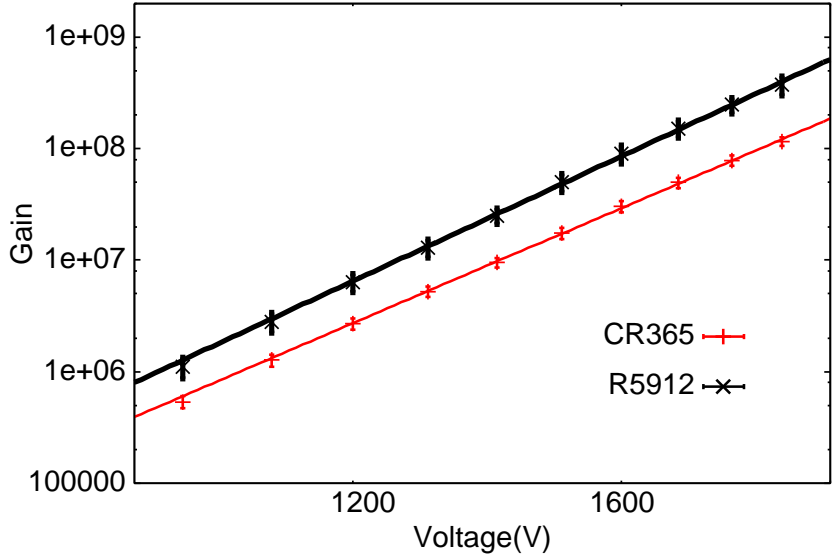}
\caption{\label{fig6} The absolute gain for CR365 and R5912 as a function of high voltage.
The gain of R5912 is about three times of that CR365 at 1800V.}
\end{center}
\end{figure}
\subsection{Dark count rate}

The dark count rate is the rate at which signals above a certain threshold are observed in a PMT while no light incident on the
photocathode. There is a correlation between dark pulse rate and the lifetime of a PMT, and the smaller noise of a PMT, the longer its lifetime becomes.
To check the dark noise, the data of R5912 and CR365 have been compared. Of course, to know the magnitude of the 1/3 PE threshold for a PMT, the absolute gain of the PMT must be known. The anode dark rate after 2 hours in dark at a gain of $\sim$1.5$\times$$10^{7}$ are both less than 3k Hz when the thresholds are set at 1/3 photoelectron.
\subsection{Linearity of PMT}
In order to extend the dynamic range of MD from 100 PEs to $10^{6}$ PEs, the linearity of the PMT output signals is essentially important.
In this work, the linearity of R5912 and CR365 is measured by using laser light source and filters as shown in Fig.2.
The light intensity is varied by use of different neutral density transmission filters and the corresponding response
is fitted with a function y=a$x^{b}$ where x is the relative transmission.
Fig.7 shows the linear behavior for CR365 at 1500V and R5912 at 1350V as a function of input light
corresponding to $\sim$ 1.5$\times$$10^{7}$ gain.
The points show the experimental data. The lines are fitting lines.
The fitting slope of CR365 is b=1.05$\pm$0.01 and the result of R5912 is b=1.03$\pm$0.01.
In this paper, the maximum linear current is defined as the peak current where
the deviation from the ideal linear current reaches $\pm$5\%.
From this figure, we can find that a peak linear current of CR365 is approximately 100 mA and R5912 almost achieved about 130 mA.
In order to compare the characteristics of the 8-inch PMTs obviously, we selected two PMTs with a relatively high gain in the above tests.
However, if we want to measure the number of photoelectrons up to $10^{6}$, we will select proper PMTs with a lower gain in the following tests,
as discussed in the next section.
\begin{figure}
\begin{center}
 \includegraphics[width=0.7\textwidth]{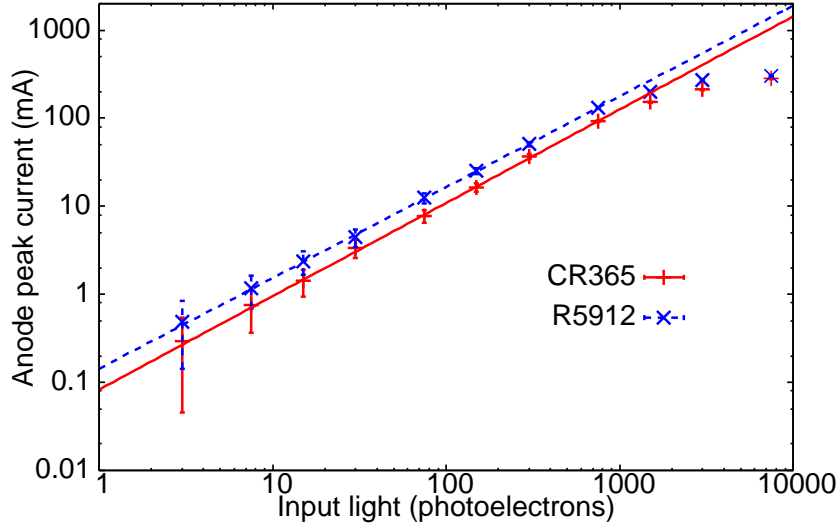}
\caption{\label{fig7} Anode peak current versus input light for the CR365-PMT at 1500V and R5912-PMT at 1350V corresponding to a $\sim$1.5$\times$$10^{7}$ gain.
The points show the experimental data. The lines are fitting lines where b is 1.05$\pm$0.01 for CR365 and 1.03$\pm$0.01 for R5912,
respectively.}
\end{center}
\end{figure}
\subsection{Extend the dynamic range of MD}

Monte Carlo simulations indicate that the highest energy showers may give rise to peak
signals as large as $10^{6}$ PEs. The PMTs, therefore, must be linear up to such
large signals. In order to extend the dynamic range of MD from 100 PEs to $10^{6}$ PEs,
we are planning to add two 8-inch PMTs in each MD cell which are responsible for the range of 100 - 10000 PEs
and 2000 - 1000000 PEs, respectively. First of all, we selected a CR365-A PMT with a $\sim$$10^{6}$ gain at 1500V to ensure
the linearity in the dynamic range of 100 - 10000 PEs. Then, with the increase of amount of incident light,
we use a 1\% filter to reduce the the input light of CR365-B  which also has a $\sim$$10^{6}$ gain at 1500V
 to realize a linear measurement of 2000 - 1000000 PEs.
Fig.8 is the output signal of two CR365-PMTs versus input light which can meet our requirements.
That is, the MD is expected to have the wide dynamic range in 6 orders of magnitude.
\begin{figure}
\begin{center}
 \includegraphics[width=0.7\textwidth]{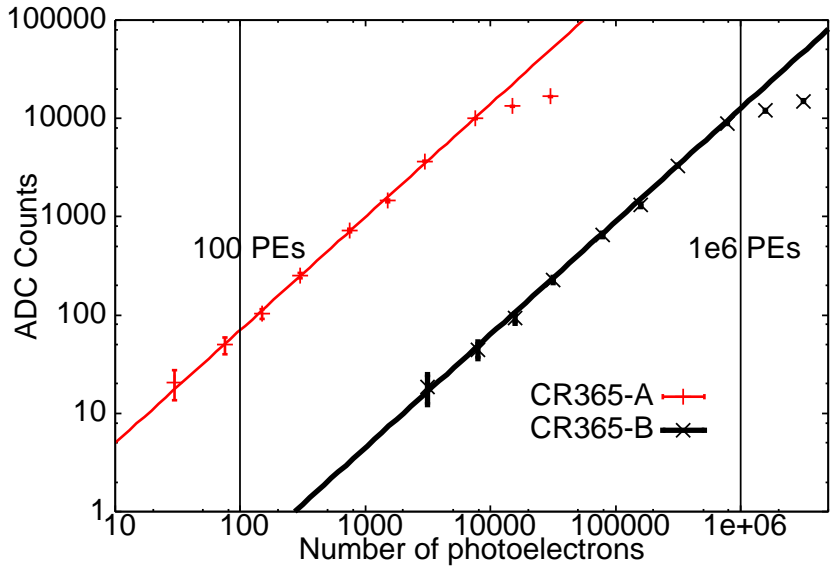}
\caption{\label{fig8} Measurement of dynamic range of CR365-A PMT and CR365-B PMT which are responsible for the range
of 100 - 10000 PEs and 2000 - 1000000 PEs. The points show the experimental data. The solid lines are fitting lines.}
\end{center}
\end{figure}
\section{Conclusion}
In this paper, we compared the characteristic features between PMT R5912 and PMT CR365. Our test experiments confirmed
that there exists no serious difference between
PMT CR365 and PMT R5912, and CR365 is within the specifications given by Beijing Hamamatsu.
This is the first comparison between R5912 and CR365. We then tested the linearity of the PMT
response by varying the light intensities and found that they satisfy our requirements
with respect to the MD having the wide dynamic range in 6 orders of magnitude.
In the near future, we will test the performance stability of this type of PMT.

\section*{Acknowledgments}
The authors would like to express their thanks to the members of the Tibet AS$\gamma$ collaboration for the fruitful discussion. This work is supported by the Grants from the National Natural Science Foundation of China (11533007, 11078002 and 11275212) and the Chinese Academy of Sciences (H9291450S3, 2013T2J0006) and the Key Laboratory of Particle Astrophysics, Institute of High Energy Physics, CAS. The Knowledge Innovation Fund (H95451D0U2 and H8515530U1) of IHEP, China also provide support to this study.\\
\appendix

\clearpage


\begin{thebibliography}{}

\bibitem{1} Kulikov and Khristiansen, JETP, {\bf 35}: 635, (1958).

\bibitem{2} M. Amenomori et al., ApJ, {\bf 678}: 1165, (2008).

\bibitem{3} J.R. H\"{o}randel, Astropart. Phys., {\bf 19}: 193, (2003).

\bibitem{4} J.R. H\"{o}randel, Astropart. Phys., {\bf 21}: 241, (2004).

\bibitem{5} Berezhko, E. G., \& Ksenofontov, L. G, J. Exp. Theor. Phys., {\bf 89}: 391 (1999).

\bibitem{6} Erlykin, A. D., \& Wolfendale, A. W., Astropart. Phys., {\bf 23}: 1 (2005).

\bibitem{7} Ptuskin, V. S., et al., A\&A, {\bf 268}: 726 (1993).

\bibitem{8} Wigmans, R., Astropart. Phys., {\bf 19}: 379 (2003).

\bibitem{9} Nikolsky, S. I., \& Romachin, V. A., Phys. At. Nuclei, {\bf 63}: 1799, (2000).

\bibitem{10} M. Amenomori et al., Phys. Lett. B, {\bf 632}: 58, (2006).

\bibitem{11} M. Shibata, Y. Katayose, J.Huang and D. Chen, ApJ, {\bf 716}: 1076-1083, (2010).

\bibitem{12} J. Huang, L.M. Zhai, D. Chen, et al., Astropart. Phys., {\bf 66}: 18-30, (2015).

\bibitem{13} LIU Jin-Sheng, et al., Chinese Physics C, Vol. {\bf 39}: No.8, 086004 (2015).

\bibitem{14} L.M. Zhai, et al., J. Phys. G: Nucl. Part. Phys., {\bf 42}: 045201 (2015).

\bibitem{15} E.H. Bellamy et al. Nuclear Instruments and Methods in Physics Research A, {\bf 339}: 468-476 (1994).

\end{thebibliography}
\bibliographystyle{JHEP}

\end{document}